\documentclass[twocolumn,showpacs,preprintnumbers,amsmath,amssymb]{revtex4}
\makeatletter
\def\@dotsep{4.5}
\makeatother

\def\ds		{\displaystyle}
\def\d		{{\rm d}}

\def\Peq	{P^{\rm eq}}
\def\Pmic	{p^{\rm mic}_{m \alpha  \beta}}
\def\Pmicp	{p^{\rm mic}_{m' \alpha' \beta'}}

\def\Nc		{C_m}

\def\Asa	{A_m}
\def\Nconf	{N_{\rm conf}}

\def\Ess	{\varepsilon_{\rm bk}}
\def\gss	{g_{\rm bk}}
\def\EH		{\varepsilon_{\rm sh}}
\def\gH		{g_{\rm sh}}
\def\Hmic	{{\cal H}^{\rm mic}_{m  \alpha \beta}}
\def\Hsig	{{\cal H}^{\rm mac}_{m  \sigma}}
\def\Hsigp	{{\cal H}^{\rm mac}_{m' \sigma'}}

\def\Hmicp	{{\cal H}^{\rm mic}_{m'  \alpha'   \beta'}}

\def\Hch	{{\cal H}^{\rm ch}_m}
\def\Hbulk	{{\cal H}^{\rm bulk}_{m \alpha}}
\def\Hps	{{\cal H}^{\rm shell}_{m \beta}}

\def\Q		{{\cal P}}
\def\Qmic	{{\cal P}^{\rm mac}_{m \sigma}}
\def\Qmicp	{{\cal P}^{\rm mac}_{m' \sigma'}}

\def\taumic	{\tau^{\rm mic}_{m,m'}}

\usepackage{graphicx}
\usepackage{dcolumn}
\usepackage{bm}

\begin{document}

\preprint{}

\title{
{How does the first water shell fold  proteins so fast~?
}}

\author{Olivier Collet}
\affiliation{
Institut Jean Lamour, D\'epartement 1, CNRS, Nancy-Universit\'e, UPV-Metz, \\
Boulevard des Aiguillettes BP 239, F-54506 Vandoeuvre-l\`{e}s-Nancy
  }

\date{\today}

\begin{abstract}
First shells of hydration and bulk solvent plays a crucial role in the
folding of proteins.
Here, the role of water in the dynamics of proteins has been investigated 
using a theoretical protein-solvent model and a statistical physics approach.
We formulate a hydration model where the hydrogen bonds between water molecules  
pertaining to the first shell of the protein conformation may be either mainly formed or broken.
At thermal equilibrium, hydrogen bonds are formed at low temperature
 and are broken at high temperature.
To explore the solvent effect, we follow the folding of a large sampling
of protein chains, using a master-equation evolution.
The dynamics shows a clear mechanism. 
Above the glass-transition temperature, a large ratio of chains
fold very rapidly into the native structure irrespective of the temperature,
 following pathways of high transition
rates through structures surrounded by the solvent with broken hydrogen bonds.
Although these states have an infinitesimal  probability, 
they act as strong dynamical attractors and 
fast folding proceeds along these routes 
rather than pathways with small transition rates between configurations
of much higher equilibrium probabilities.
At a given low temperature, a broad jump in the 
folding times is observed. 
Below this glass temperature, the pathways where hydrogen bonds 
are mainly formed
become those of highest rates although with conformational changes of huge relaxation times.
The present results reveal that folding obeys a double-funnel mechanism.
\end{abstract}


\maketitle

To this date, the three-stranded $\beta$ sheet is the faster folder
finding its native structure in the amazingly short time of 
140 nano-seconds~\cite{Xu2006}.
The protein folding problem is still considered as one of the major
unsolved problems of science~\cite{Science2005} and the answer to the Levinthal
question~\cite{Levinthal1968} "how a protein can fold so fast~?"
remains a "grand challenge"~\cite{Dill2007}.


Protein folding is the process whereby a protein folds into its native 
structure. The slowest folding proteins may require a few minutes due among other factors
to proline isomerization \cite{Kim1990}. They fold passing through many intermediate states.
On the other hand, many small single-domain proteins fold very rapidly over 
time scales of
a few microseconds \cite{Jackson1998,Kubelka2004}. 
For many of these proteins, the folding process is a single exponential function of 
 time \cite{Jackson1998,Kim1990} and is modeled by a two-state mass action model
and an Arrhenius diagram 
on which the free energy of some ensembles of chain conformations is plotted as a 
function of reaction coordinate, usually not known.
This diagram exhibits a transition state between the unfolded and the
native states \cite{Schonbrun2003}.

Moreover, some ultra fast folders exhibit more complex kinetics with 
non-Arrhenius behavior \cite{Ghosh2007} (i.e. non-linear dependence of
the logarithm of the folding rate on the inverse of the temperature).
Some results show that the activation energy is positive 
at room temperature, decreases as the temperature increases and may become
 negative at high temperature \cite{Zhu2003}.
It has been suggested that this could arise from the temperature dependence 
of the hydrophobic effect \cite{Chan1998b,Akmal2004}.


An alternative to this transition-state view is the concept of folding funnel \cite{Leopold1992}.
This energy-landscape picture is based on the idea of minimal frustration \cite{Bryngelson1995},
which states that the evolutionary mechanism has retained those protein sequences that
have a funnel-like energy landscape. In that concept, 
the height of the funnel represents the conformational energy and 
its width  represents the entropy of the subset of chain conformations of a given energy
\cite{Onuchic1996,Wolynes1997,Socci1998,Onuchic2004,Clark2004}. The top of the funnel
is populated by the huge number of denaturated configurations with a large energy and
entropy and the bottom with the unique native structure of very low energy and quasi-nil
entropy. 
Each protein chains folds from the top of the funnel towards the bottom.


The transition-state theory and the folding-funnel picture are two different approaches.
The first one describes well the two-state kinetics, but does not explain why folding is
so fast.
The second one explains well why folding is so fast, 
and  the thermodynamic free energy barrier, that gives rise to two-state kinetics
and makes transition-state theory applicable to the folding process, is essentially of entropic nature.



Lattice models of proteins are among the favorite tools for the theoretical study of folding.
The microscopic representation of the proteins is  simplified to allow large sampling
of the configurational space. Proteins are modeled as self-avoiding-walk chains of beads,
which are located on a two or three-dimensional square or cubic lattice. 
For small-length chains, a full enumeration of the conformations allows the
exact calculation of the partition function, and, thus of statistical averages
\cite{Lau1989,Go1983,Dinner1994,Collet2001}.
The Hamiltonian of a given conformation of the chain results from the interaction of the
first neighbors of the beads on the lattice, but not in the sequence. 
The more popular model of 
couplings between monomers are known as the G\~{o}\cite{Go1983}, 
the HP \cite{Lau1989,Chan1991,Chan1994,Chan1998} and the random-energy model(REM) \cite{Derrida1981,Shakhnovich1990a,Sali1994a,Sali1994b,Gutin1995b,Gutin1995}.
In the G\~{o} model, the interaction between the beads of a given compact structure, chosen as the native conformation, is set to -1 and all the other couplings to 0. In the HP model, 
the sequence of the protein is given in terms of a series of hydrophobic (H) or polar (P) residues and the coupling
between two hydrophobic beads is set to -1 and that involving at least one polar monomers to 0.
In the REM, a matrix of couplings between all pairs of residues is constructed by
drawing random numbers from a Gaussian distribution. 


Despite the numerous results obtained with such models, they fail to
reproduce a fundamental feature of a protein~: its cold denaturation.
This mechanism is associated to the loss of stability of the native structure
upon cooling down the system \cite{Privalov1990,Nishi1994,Kumar2006}.
For half a century, it has been well known that water plays a crucial role
in the mechanism of folding \cite{Kauzmann1959,Warshel1970,Dill1990}
and an understanding of cold denaturation requires a finer model of the
solvent than the temperature-independent attractive parameters used in G\~{o}, HP or REM.
Some recent refinements of the models, including temperature dependence of 
the hydrophobic effect, have allowed
to model the cold transition \cite{DeLosRios2000,DeLosRios2001,Collet2001,Collet2005} 
to be described.


As the physics of protein folding proceeds in water and the interactions of the 
protein chemical groups are solvent mediated, the representation of the solvent is of
great importance \cite{Chan2004}.
The hydrophobic effect is an active field of research by itself \cite{Garde1996,Hummer1998a,Garde1999,Gomez1999,Lyndell1997}.
It represents the tendency of water to exclude nonpolar solutes.
It results from a disruption of the network of hydrogen bonds between water molecules
caused by the transfer of an nonpolar solute into water.
The energy variation of this process is favorable at room temperature, whereas 
the entropy cost leads to a large positive free energy of transfer.
In addition, the physics of the solvent-mediated interactions of the protein may be
captured by studying the interaction
of two nonpolar solutes immersed in water \cite{Chan2004}. This
prototypical interaction can be handled by averaging e.g. the
degrees of freedom of the solvent in the free-energy function of two methane solutes 
in explicit water \cite{Pratt1977,Shimizu2000}.
The profile of the free energy as a function of the distance between the
two methane solutes shows a deep minimum for the contact distance between the
two molecules and another one, less pronounced  when they are separated 
by a distance slightly smaller than one solvent molecule diameter. 
A maximum, higher than the free energy of the 
two isolated solutes arises between these two configurations. 
This maximum  is known as the desolvation barrier.
As the temperature increases,
the contact between the nonpolar solutes becomes more favorable as the desolvation
barrier is reduced.
This barrier tends to favor a high thermodynamic cooperativity of the model in contrast with
a model without desolvation barrier \cite{Kaya2003,Kaya2005}.
It has also been shown that the physics behind this barrier is responsible for 
the large diversity in the folding rates, similar to what is observed experimentally \cite{Ferguson2009}
%

Moreover, recent experimental work has shown that structural fluctuations of the solvent 
may control structural fluctuations of the protein \cite{Fenimore2004, Frauenfelder2006,Lubchenko2005,Shenogina2008,Frauenfelder2009}.
It has also been observed that motion of hydration water drives protein dynamics \cite{Zanotti2008}.
This could be responsible for the protein-solvent dynamical transition connected with the liquid glass transition of hydration water \cite{Doster1989}.
These results show the importance of the degrees of freedom of the hydration first shell
for the dynamics of the proteins.
%
A theoretical model of the hydrophobic effect introduced by Muller \cite{Muller1990}
and extended by Lee and Graziano \cite{Lee1996} allows to separate the contribution
of the hydrogen bonds of the first shell and that of the bulk
water.
The basic idea stems from the result that the hydration of nonpolar solutes 
presents a large entropy cost and a small favorable energy.
The hydrogen bond breakage in the bulk is considered as
a two-state equilibrium between the formed and the broken hydrogen bonds.
The equilibrium constant between the two states is related to the fraction
of formed hydrogen bonds 
and to the difference in  enthalpy and the ratio of degeneracy resulting
from the breaking of one hydrogen bond.
A similar description is used for to the water molecules of the first shell.
It considers that the thermodynamics of a broken hydrogen bond in the first 
shell is the same than that in the bulk and gives a picture of the hydrophobic effect
based on the enthalpy gain and entropy cost arising from the creation of a bond
in both situations.
A little bit later, Lee and Graziano \cite{Lee1996} pointed out that the energy associated to 
a broken hydrogen bond is not the same for a water molecule of the first shell 
and for one of the bulk.
The presence of an nonpolar solute induces the breakage of a hydrogen bond
of the first shell, leading to a more unfavorable energy than the same event in the bulk.
The two-state model of the hydrophobic effect has been applied \cite{Silverstein1999} to the 
two-dimensional Mercedes-Benz model of water \cite{BenNaim1970,Becker2006}.
As a result, they give a spectrum where the non-degenerated ground state is for the formed hydrogen
bond in the first shell and the highly degenerated states for the broken hydrogen bond 
in the first shell corresponds to the larger value of the spectrum. 
The energies and the degeneracies of the formed or broken bonds
in the bulk are found between the two previously described.

In this paper, this picture has been reduced further by gathering together the two close 
energy levels associated to the broken and formed hydrogen bonds of bulk water
\cite{DeLosRios2000} and has been applied to a lattice model of protein
to study the effect of the first shell on the protein dynamics.
Aside from the hydrophobic model itself, how the solvent is simulated has a significant impact on the energy landscape of the protein.
Explicit solvent models are very computationally expensive \cite{Chipot1999}. 
Implicit solvent models have been developed to take into account the solvent as a mean field effect \cite{Lee1971,Eisenberg1986,Wesson1992,Ooi1993,Collet1996a,Premilat1997,Collet2008}.
Yet, results obtained from explicit simulations do not always agree with those from implicit models \cite{Collet1997a,Liu2005}.
Up until now, the strategy to follow the kinetics of the proteins consisted in
averaging the degrees of freedom of the solvent by calculating the
free energy of solvation of each protein structure and the transition rates between two protein conformations.
The system evolves along effective routes made of conformations 
surrounded by an averaged solvent.

Here, as the solvent model allows it, we have 
 {\it 
calculated  the rate between two protein-solvent microscopic configurations and
 grouped together some equivalent transitions. }
The physical pathways are microscopic routes in the protein and 
the solvent configurational space,
not "mean" routes in the conformational space of the protein surrounded by an effective solvent.

The dynamics of a large set of chains in the solvent is calculated using a master equation evolution.
In the spirit of the concept of folding funnel \cite{Leopold1992,Onuchic1996,Wolynes1997,Onuchic2004},
a picture of the folding in terms of two surfaces, depending on the state of the hydrogen bonds of the first shell solvent,  drawn in an entropy-energy plot, is given.
The mechanisms responsible for the fast folding and the glass transition 
are detailed in the body of the paper.
In the first part, the protein and the solvent models are described. 
In the second part, the equations of the evolution are established.
In the third part, the mechanism, which occurs during the fast overcoming of
kinetic barriers is explained, then
the mechanism responsible for the glass transition is revealed.

\vskip0.5cm

\section{Model.}

The microscopic Hamiltonian of the chain in conformation $m$, surrounded by
a first shell of solvent molecules in configuration $\beta$ and bulk 
solvent molecules in structure $\alpha$ is denoted~:
$$\Hmic = \Hch + \Hps + \Hbulk$$
The first term results from the intrachain interactions,
the second one from the contribution of the molecules of the first shell 
solvent in interaction with the protein 
and the last one from that of the bulk water.

\subsection{Protein Model}
The proteins are represented as self-avoiding walk strings of $N$ monomer beads located on a two-dimensional lattice \cite{Lau1989,Gutin1995,Dinner1994} (here $N=12$). 
This length of the chain is short enough
to allow analytical calculations for the dynamics and long enough to give interesting
results.
The compactness of a structure $m$ is the number of intrachain contacts of the chain conformation $m$~:
$\Nc = \sum_{i \ge j+3} \Delta_{ij}^{(m)}$
where $\Delta_{ij}^{(m)}=1$ if the monomers $i$ and $j$ are first neighbors 
on the lattice and 0 otherwise.
The accessible surface area of the conformation $m$ to the solvent is defined 
as the number of links between the chain beads and the empty sites of the lattice~:
$\Asa = 2N+2-2\Nc$.
The intrachain Hamiltonian of the peptide structure $m$ is~:
$$\Hch = \sum_{i \ge j+3} B_{ij} \ \Delta_{ij}^{(m)} $$
To model the heterogeneity of the sequence of amino-acids of the chain,
the couplings $B_{ij}$ between monomers $i$ and $j$ are drawn at random
from a Gaussian distribution centered on $-2$ with standard deviation equal to 1 \cite{Shakhnovich1990a}.
Such a way of designing the sequences leads to create a configurational
space with small energy gaps between the structures of bottom of the
energy spectrum. A particular compact (native) structure does not emerge as a stable
conformation of the sequence with a large energy gap with other compact conformations. 
To increase the stability of the native conformation of the sequence\cite{Hardin2002a,Hardin2002b},
we select a compact conformation for the native structure of the sequence and
we rank the couplings such that the
minimum ones are associated with the native contacts.

\subsection{Solvent Model}
For each chain conformation, 
the empty nodes of the lattice models the solvent effect\cite{Collet2001,DeLosRios2000}.
We do not attempt to introduce a fine description of the structural properties of 
solvent around proteins itself, but we describe a realistic solvent effect
on the weights of the chain conformations.

The bulk water contribution is simply modelled by an extensive negative free energy term
which guides chains towards compact structures.
The microscopic structures of the first shell solvent around any
given chain conformation are separated into two groups 
depending on whether most of the hydrogen bonds are formed or not. 
Hence, each protein conformation has two possible values for its energy depending
on the structure of the first shell:
one for a ground state (GS) associated to a rather organized first shell 
and another one for an excited state (ES) with mainly broken hydrogen bonds.
\begin{figure}[hbtp]
\includegraphics[width=8cm]{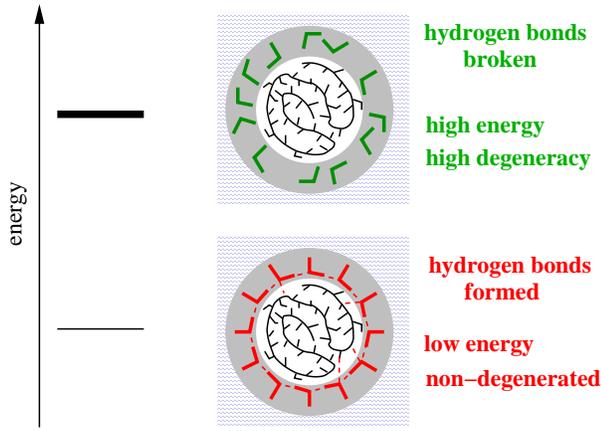}
\caption{\label{solvent_model}
The solvent around a chain conformation chosen at random.
The unique solvent configuration where the hydrogen bonds between water molecules and between 
water and the chain are formed (shown in the left picture) is the non-degenerated ground state (GS) of the first
shell and the other structures (one is shown in the right picture), where the hydrogen bonds are mostly broken, 
are grouped together in the highly degenerated excited state (ES).
Only, the highly organized and highly disordered solvent configuration are taken into
account.
All the other cases are not
considered in this two-state picture which only
takes into account the lowest energy and largest entropy macro-states
which are the most important contribution for the statistical physics approach.
}
\end{figure}

For each chain conformation $m$, the first shell interaction is  
extensive with respect to $\Asa$.
The links between a solvent node and the chain beads account for the first shell contribution. 
The non-degenerated ground state denoted by $\beta =0$ models the
first shell water molecules with formed hydrogen bonds around the protein. 
It is taken as the energy reference which equals 0.
The excited states, corresponding to $\beta \ge 1$, are for the $\gH^{\Asa}$ 
first shell structures with broken hydrogen bonds. Their energy is $\Asa \ \EH$.
The Hamiltonian of the first shell solvent for the chain structure $m$ is written~:
$$
\Hps = \Asa \EH \sigma(\beta)
\ \ \  {\rm with} \  
\left\{
\begin{array}{l}
\sigma(0) = 0 \\
\sigma(\beta)=1 \ {\rm if} \ 1 \le \beta \le \gH^{\Asa} 
\end{array}
\right.
$$

When one intrachain contact is formed, two monomers-solvent bonds are broken.
Then, after removing the constant term, the pure solvent contribution
is a simple function of $2 \Nc$\cite{Collet2001,Collet2005}.
The factor of 2  guarantees that the solvent volume does not
depend on the chain structure.
For each chain structure $m$, the $\gss^{2 \Nc}$-fold degenerated Hamiltonian 
of the pure solvent is independent of the bulk micro-state~: 
$$ \Hbulk = 2 \Nc \Ess \qquad {\rm for} \ 1 \le \alpha \le  \gss^{2 \Nc}$$

The results given in the paper hold while the parameters are ranked as
follow~: $0 < \Ess < \EH$ and $\gss < \gH$.
In the spirit of the results obtained by Silverstein  {\it et al}\cite{Silverstein1999},
the values of the solvent parameters are ranked as follow 
$\Ess=0.2$, $\EH=0.6$, $\gss=2$ and $\gH=3$.  
The reported results in this paper are quite robust with regards to change
the parameters holding the above ranking equation.
However, for some technical reasons (discussed below) if $\gss$ or/and $\gH$ are chosen
too large, the computational times of the calculation becomes too prohibitive.

The energy and the degeneracy of the ground and excited macro-state of each peptide conformation are~:
$$\label{hms} 
\left.
\begin{array}{l}
\Hsig = \Hch + 2 \Nc \Ess + \sigma \Asa \EH \\
\ds g_{m\sigma}  = \gH^{\sigma  \Asa} \ \gss^{2 \Nc} \nonumber
\end{array}
\right\}
\ {\rm with} \ \sigma = 0 \ {\rm or} \ 1
$$

\subsection{Results for the chain in interaction with the solvent}

\begin{figure}[hbtp]
\includegraphics[width=8.3cm]{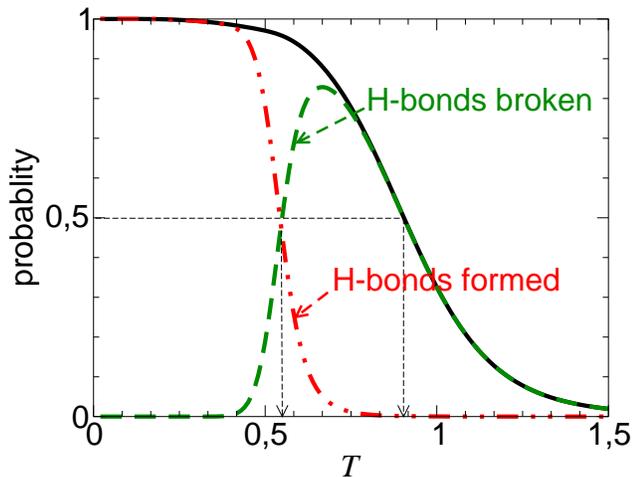}
\caption{\label{main_result_equil}
Equilibrium probabilities of occurrence of the native structure (black solid line)
as functions of the temperature (given in arbitrary units) with
the contribution of the ground state (red long dashed line) 
and of the excited state (green dashed line).
At a temperature $T_0$, the probability that the hydrogen
bonds are broken becomes significant and at $T^*$, 
the probability to observe hydrogen bonds around the native conformation
is the same as that to observe water with broken hydrogen bonds.
The probability of occurrence of the native structure equals 1/2 at $T_m$.
}
\end{figure}

The thermal equilibrium probability of each macro-state is given by~:
${\cal P}_{m \sigma}^{\rm eq} = g_{m\sigma} \exp(-\Hsig/T) / Z(T)
$ and that of a chain structure is $\Peq_m = \sum_{\sigma=0}^1 {\cal P}_{m \sigma}^{\rm eq}$.
Here the Boltzmann constant is set to 1 and 
the partition function is:
\begin{eqnarray}
Z(T) & = &\ds \sum_m 
		\sum_{\alpha=1}^{\gss^{2 \Nc}}
		\sum_{\beta=0}^{ \gH^{\Asa}} 
	\exp\left(- \frac{\Hmic}{T} \right)      \nonumber \\
	& = & \ds \sum_m \gss^{2 \Nc} \exp \left(- \frac{\Hch+\Hbulk}{T} \right) 
\nonumber \\
\label{Z}	
&    & \ds 
\left[	1 + \gH^{A_m} \ \exp \left(-\frac{ A_m \EH} {T}\right) \right]  
\end{eqnarray}
The native conformation is the structure of largest value of 
$P_m^{\rm eq} $
determined by a full enumeration of the conformational space of the chain at low 
temperature. 
For the set of couplings $B_{ij}$ and solvent parameters used here,
the native conformation
is a compact structure of intrachain energy -9.895. 
The folding transition temperature is defined as, 
the melting temperature $T_m$ of the experimental literature\cite{Privalov2000,Faisca2006}
 at which the equilibrium probability of the
native structure equals that of all the other denatured conformations 
One specific chain structure (the native conformation) has a larger equilibrium
probability to occur than all the other conformations for $T<T_m$ or in others words $P^{\rm eq}_{\rm Nat}(T_m) = 0.5$
(here $T_m=0.90$).
Figure \ref{main_result_equil} shows that the equilibrium probability of occurrence of this
native structure (Nat) surrounded by water with formed hydrogen bonds reaches one
at low temperature. 
Other solvent configurations around Nat become relevant for $T > T_0$ (here $T_0=0.45$).
Last, the probability of occurence  of the native structure with of the first shell
solvent in ES equals that in GS at a specific temperature denoted $T^*$
(here $T^*=0.54$).

We note in passing, as the solvent parameter of the GS of the first shell is lower than that of the bulk,
it could be possible to observe cold denaturation of the chains, if the GS of the
extended chains would be the state of lowest energy among the whole configurational space \cite{Collet2001,Collet2005}.
Here, however, the parameterization chosen avoids this possibility in order to study only the folding mechanism.

\begin{figure}[hbtp]
\includegraphics[width=7.0cm]{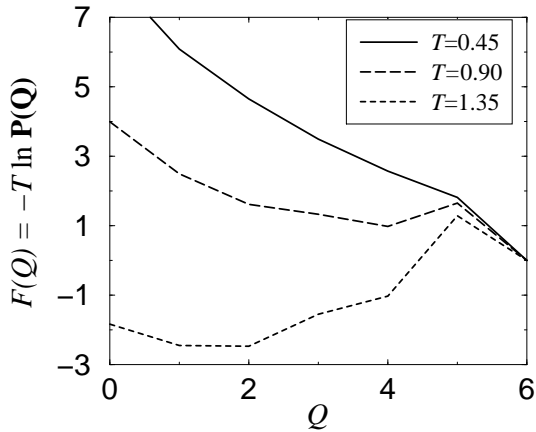}
\caption{\label{fq}
Free energy profile for the lattice model at the three 
temperatures indicated. $P(Q)$ is the probability of occurrence
over $Q$ at thermal equilibrium.
}
\end{figure}

Figure \ref{fq} shows the conformational distribution as function of the number of 
native intrachain contacts calculated as follow~:
$$F_T(Q) = - T \ln \sum_m \delta(Q-Q_m)  \sum_\sigma g_{m \sigma} \exp(-\Hsig/T)$$
where $\delta(0=1$ and 0 otherwise and $Q_m$ is the number of native contact of
the chain structure $m$.
Under denatured conditions ($T > T_m$), the free energy profile show a barrier between
the native and non-native conditions. The sets of conformation with one or two
intrachain contacts, surrounded by solvent in ES (see fig. \ref{cv}) have the largest
probability of occurrence.
At the melting temperature, the same barrier still separates the equiprobable native 
and non-native populations.
Under strong native condition (low temperature), the shape of the free energy profile
is similar to that observe for a downhill folding.

Indeed, as the temperature decreases, the more probable non-native population 
shifts from the sets with few native contacts to that with the maximal native contacts

\subsection{Results for the solvent at equilibrium}

At thermal equilibrium with a bath at temperature $T$, the 
probability of occurrence of the conformation 
$m$ with the solvent in micro-state $(\alpha \beta)$
tend towards~:
$p^{\rm eq}_{m \alpha \beta} =\exp\left(-{\Hmic}/{T} \right) / Z(T)$.

\begin{figure}[hbtp]
\vskip0.4cm
\centerline{\includegraphics[width=8cm]{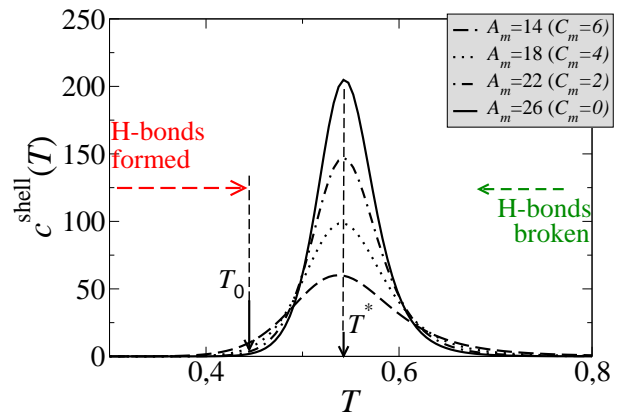}}
\caption{\label{cv}
The heat capacity of the solvent around the chain structures depends on the
compactness of the conformation. At low temperature, the chain are surrounded
by rigid cages of water molecules. The higher the compactness, the higher is the
temperature of occurrence of the chain with broken hydrogen bonds of the solvent.
}
\end{figure}

The mean energy of the first shell solvent around chain conformation $m$ is~:
$$U_m^{\rm shell}(T) 
= \frac{\sum_{\alpha \beta}p^{\rm eq}_{m \alpha \beta} \Hps } 
  { \sum_{\alpha \beta}p^{\rm eq}_{m \alpha \beta} }
= \frac{\Asa \EH \gH^{\Asa} \exp(-\Asa \EH / T) }
{1 + \gH^{\Asa} \exp(-\Asa \EH / T) }
$$
and the heat capacity is~:
$$c_m^{\rm shell}(T) = \frac{\d U_m^{\rm shell}}{\d T} = 
\frac
{{\Asa}^2 \EH^2 \gH^{\Asa} \exp(-\Asa \EH / T) }
{T^2 [1 + \gH^{\Asa} \exp(-\Asa \EH / T)]^2 }
$$
These curves as function of the temperature only depends on the chain
exposure to the solvent, {\it i.e.} on the compactness (fig.\ref{cv}).
They exhibit a maximum at the same temperature: $T^* = 0.54$.
It is the temperature of equiprobability of occurrence of the two phases: 
broken and formed hydrogen bonds of the first shell around the peptide.
For $T_0 < T < T^*$, the solvent configurations with a water cage around the 
peptides is preferred  to the broken hydrogen bonds. For $T < T_0$, it is the 
only relevant state. Above $T^*$, the solvent occurs in the excited macro-state.
This is in good agreement with the result obtained for the thermodynamics of 
the chains presented above.

\section{Time evolution.}

To explore all the possible routes
 from the non-native structures to the native conformation,
the probabilities of each micro-state, composed of one protein structure in interaction 
with one solvent configuration,
evolve using a continuous time Markov process applied to a large sampling of peptides.

Master equation approach to protein folding has already be used in 
lattice model with an effective solvent \cite{Cieplak1999}.
It is shown in appendix A, that a master equation of the macro-states may be
deduced from the master equation of the micro-states. 
In a finite time approach, the probabilities of the macro-states evolve following the Euler algorithm\cite{Cieplak1998}~:
$$\Qmic(t + \delta t) = \Qmic(t) + \delta t \sum_{m'} \sum_{\sigma'}  
Y_{m \sigma, m' \sigma'} \Qmicp(t)$$
where 
$$Y_{m \sigma, m' \sigma'} = 
g_{m \sigma}\  \frac{V^{(0)}_{m m'}}{\taumic} 
(1+\exp((\Hsig - \Hsigp)/T)^{-1}  $$
for $m \ne m'$ or $\sigma \ne \sigma'$ and
$$Y_{m \sigma, m \sigma} = - \sum_{m' \sigma' \ne m \sigma} Y_{m' \sigma', m \sigma}$$
As explained in appendix A,  
$V^{(0)}_{m m'}=1$ if structures $m$ and $m'$ are connected by a one monomer move
(either a corner flip or a tail move) or if $m=m'$.
The characteristic time associated to a chain move ($\taumic=\tau_c$ if $m \ne m'$)
and to a solvent move ($\taumic = \tau_s$ if $m=m'$) are set to 
$\tau_c = 1$ and $\tau_s = 0.001 \ll \tau_c$.

\begin{figure}[hbtp]
\centerline{\hskip0.6cm\includegraphics[width=7.cm]{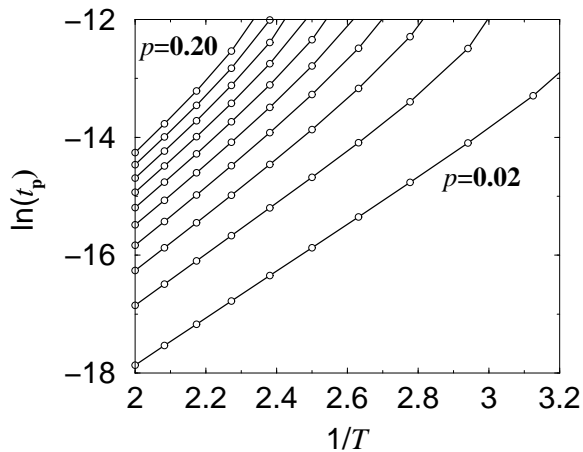}}
\vskip0.7cm
\caption{\label{main_result_tfold}
Logarithm of the  waiting time $t_p$ to observe the native structure with a probability equal
to $p$ plotted as a function of the inverse of the temperature for $p=0.02$ to $p=0.20$ by steps
of 0.02.
}
\end{figure}

A sufficient condition to conserve the norm of the probability vector 
(i.e. $\sum_{m \sigma} \Qmic(t) = 1$) is satisfied by fixing 
$\delta t = 1 / \max_{m \sigma} \{ Y_{m \sigma,m \sigma} \}$.
As the values of $g_m\sigma$ and then those of $Y_{m \sigma, m' \sigma'}$ may be huge,
the value of $\delta t$ is tiny. Then, the evolution of the probability vector is
very slow.

\begin{figure}[htbp]
\centerline{\hskip0.6cm\includegraphics[width=6.5cm]{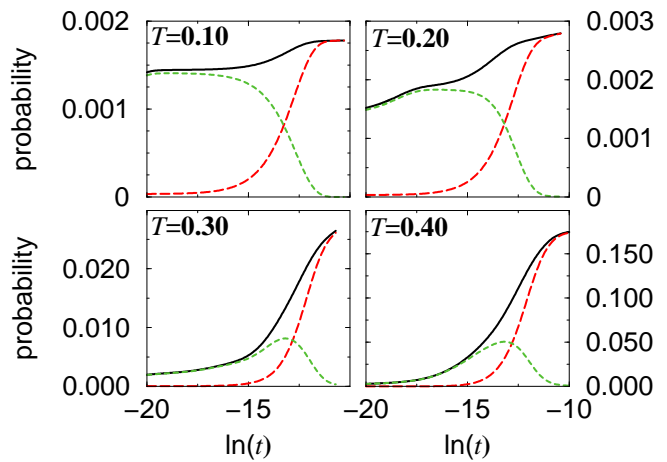}}
\vskip0.7cm
\caption{\label{main_result_kinetic}
Evolution of the probabilities of occurrence of native structure (back solid line),
the contribution of excited state (green dashed line) and of the ground state (red long dashed line)
as functions of the time at different temperatures.
  For temperature of 0.10 and 0.20,
some chains topologically close to the native
structure reach it very fast because no activation barrier occurs in their pathway.
Then, the probability of the ES decreases monotonically to zero and that of
the native structure reach a plateau and the kinetics is frozen.
As the number of chain structures of the basin do not depend 
on the temperature, the plots have very similar shapes and 
the values of the probabilities are very close to each other showing that a local
phenomenon is taking place.
  For temperature of 0.30 and 0.40,
the ES of the native conformation (which have a null equilibrium probability) 
acts as an attractor, in the early events, and reach very rapidly a  maximum
of its probability  of occurrence, much larger than its equilibrium probability.
A fast transition takes place towards the GS of the native structure.
Then, a slower dynamic regime guides the other chains towards the native structure.
Now, the values of the probabilities depend on the temperatures showing that
a global mechanism dominates. 
}
\end{figure}

The simulations of folding start with the initial condition : $\Qmic(0) = 1 / 2 \Nconf$
where $\Nconf=15019$ is the number of chain structures.
The probability of occurrence of conformation $m$ after time $t$ is
$P_m (t) = {\cal P}^{\rm mac}_{m ; 0}(t) + {\cal P}^{\rm mac}_{m ; 1}(t)$.
The waiting time to observe the native structure with a probability $p$ is noted $t_p$.
The main results of the calculation are summarized in fig.\ref{main_result_tfold}
and \ref{main_result_kinetic}.
They exhibit some findings on the out of equilibrium folding of a large
sampling of chains at different temperatures.

Figure \ref{main_result_tfold} shows a non-Arrhenius behavior of the model.
Because of the tiny value of $\delta t$, the early events of the folding are shown, here.
The remaining part of this plot will be deduced from the results given below.
It will be shown that even if the curves may be well fitted by a Vogel-Fulcher-Tamman 
function ($t_p(T) \propto \exp(-E_a / (T-T_0))$), which is the signature of an 
$\alpha$-relaxation with a dynamical temperature $T_0 = T_0(p)$, the remaining part
of the curves exhibits a more complex shape which can not be capture by effective
solvent models.

Figure \ref{main_result_kinetic} shows that, even if its equilibrium probability is infinitesimal,
the excited state of the solvent acts as a strong
dynamical attractor above a particular temperature to be discussed later. 
Below this temperature the kinetics is the same whatever the temperature.

\section{Relaxation times of the moves.}

Only two connected macro-states $(m\sigma)$ and $(m'\sigma')$ are considered for a while.
The equations \ref{Ymaster} and \ref{Yrate} show that the rate of the transition $(m'\sigma')
\rightarrow (m\sigma)$ depends on the degeneracy of the macro-state $(m\sigma)$
and not of that of $(m' \sigma')$.
In other words, the increase of the probability of $(m \sigma)$ and the decrease of
that of $(m' \sigma')$ is due to the degeneracy of $(m \sigma)$ and 
to the difference of energy between the states
with $(m' \sigma')$.
After thermal equilibration, the probability of $(m \sigma)$ would go towards
${\cal P}^{(\infty)}_{m \sigma} = 
{{\cal P}^{\rm eq}_{m \sigma} }/
({ {\cal P}^{\rm eq}_{m \sigma} +  {\cal P}^{\rm eq}_{m' \sigma'} } )
$
The solution of the system of equations \ref{Ymaster}, for only two states is~:
$${\cal P}^{\rm mic}_{m \sigma}(t) = 
{\cal P}^{(\infty)}_{m \sigma} + [{\cal P}_{m \sigma}(0) - {\cal P}^{(\infty)}_{m \sigma}] \
\exp(-t/ \overline \tau_{m \sigma, m' \sigma'} )]$$
with a very small relaxation time~: 
\begin{equation}\label{time}
\overline \tau_{m \sigma, m' \sigma'}  = 
(Y_{m \sigma, m' \sigma'} + Y_{m \sigma, m' \sigma'} )^{-1} \ll \taumic
\end{equation}
which now depends on the energies and degeneracies of the macro-states.
While numerous chains initially in state $(m \sigma)$ move into state 
$(m' \sigma')$, some of them may go back to $(m \sigma)$.
As a consequence, the relaxation times depend on forward and backward rates of a move.
As an aside, if the move from $(m \sigma)$ to $(m' \sigma')$ were allowed and not
the backward transition, then the relaxation time would simplify  to 
$ 1 / Y_{m \sigma, m' \sigma'}$.

As the  degeneracy of the excited state is always larger than that of the ground state,
the value of the rates between two excited states of the chain is larger than that
between two ground states (if the energy difference is of the same order) 
As a consequence, the relaxation times of the former transitions  are always very small
at not too low temperature. 
They are supposed to simulate peptide evolving in a fluid solvent.
In contrast, the relaxation times of the latter transitions are larger. This models chains evolving in a viscous medium.
Thus, the dynamics of the chains surrounded by water with broken hydrogen bonds 
is faster than that of low degenerated structures in interaction with
solvent with formed hydrogen bonds. 

In addition, the more extended the structure, the larger the degeneracy of the
excited states and the smaller is the relaxation time.
This is because a more extended chain has a larger exposure to the solvent and
thus a lot of hydrogen bonds are broken in the first shell and as a consequence
the dynamics is faster.

Last the connexion between the excited and ground states of the same chain 
conformation ($m \equiv m'$)
have a small relaxation time because $\tau_s \ll \tau_c$.
The difference between that case and the fluid connexion depends on the ratio 
$\tau_c / \tau_s$. 

\section{Fast folding mechanism}

\begin{figure}[htbp]
\centerline{\includegraphics[width=8.0cm]{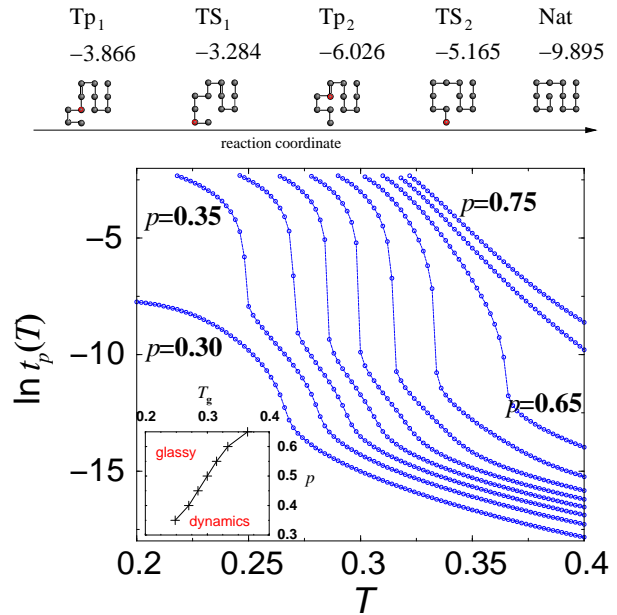}}
\caption{\label{example}
Top:  
A possible pathway with five chain structures connected by one monomer moves ending in Nat
taken as an example to capture the role play by the solvent in the folding kinetics.
The energy of the GS is given above of each conformation.
A single route from Tp$_1$ to Nat is considered, to simplify the study of this first approach, and the reaction coordinate is simply a function (not given) of the 
number of conformational changes necessary to reach Nat.
Bottom: 
the waiting times to observe a ratio $p$ of proteins in Nat, starting from
an equiprobability of all the states, show a glass transition at temperatures 
depending on $p$ (shown in the inset).
}
\end{figure}

To understand the mechanism underlying the fast folding,
a possible pathway leading to the native structure,
shown in fig.\ref{example}, is considered as a first approach of the problem.
This pathway consists in five chain structures connected by one monomer moves.
In contrast with the whole conformational space where there exist a great number of
routes from a given conformation to the native structure, in this section we first
consider a simplified trajectory where there is a unique pathway 
(via TS$_1$, Tp$_2$ and TS$_2$) from Tp$_1$ to Nat.
Structures Tp$_1$ and Tp$_2$ have five intrachain contacts and thus, smaller 
energy than TS$_1$ and TS$_2$ which have three intrachain contacts.  
Then, the GS of the structures Tp$_1$ and Tp$_2$ should act as kinetics traps
and that of TS$_1$ and TS$_2$ as transition states towards the native conformation.
At $t=0$, the initial probability of each macro-state is set to $P(t=0) = 0.1$.
An Euler algorithm is applied to simulate the evolution of the probabilities of the ten macro-states of this subsystem composed by the ES and the GS of this five 
chain structures.
The waiting time $t_p$ to observe the native structure with a probability equal
to $p$ is plotted for $p=0.30$ to $p=0.65$ by steps of 0.05 as function of
the temperature.
At $t=0$, the probability of occurrence of Nat is already 0.20 and 
for $p<0.30$, the waiting time is tiny since a many chains in the TS$_2$
states fold instantaneously in Nat.
For $T<0.20$, the waiting time to reach a ratio $p=0.30$ becomes constant, 
but that to observe a larger ratio become huge.

For $p>0.30$, the waiting times increase continuously as the temperature
is decreased until a temperature, depending on $p$ and denoted by $T_g(p)$,
where a broad dynamical transition occurs. 
The kinetics is fluid above $T_g$ and glassy below.

\begin{figure}[htbp]
\centerline{\includegraphics[width=8.0cm]{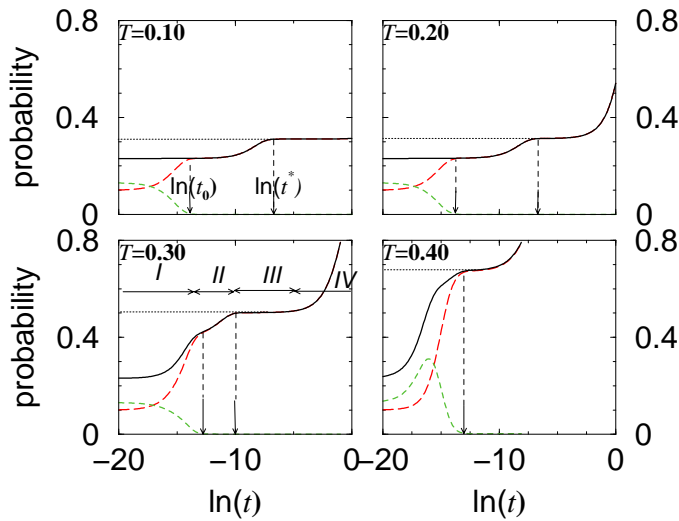}}
\vskip0.5cm\caption{\label{evolution}.
Evolution of the probabilities of the native structure (black solid lines)
and the ES (green dashed lines) and the GS (red long dashed lines) contributions 
as functions of the time for different temperatures.
}
\end{figure}

Figure \ref{evolution} shows the very early events of the folding of this
small system. 
The ES of Nat relax into the GS via a, very fast, solvent
transition. At a temperature independent time, $t_0$, 
the probability of the ES of Nat becomes very small. 

At low temperature ($T=0.10$ or $T=0.20$), half of the chains in conformation TS$_2$
fold in Nat and the other half goes to Tp$_2$ between the time
$t_0$ and  a temperature dependent time, $t^*$  after which the probability
of occurrence of the native structure becomes constant for a while. 
Then, a long plateau with $p\approx 0.3$ occurs where the dynamics is glassy.

\begin{figure}[htbp]
\centerline{\includegraphics[width=8cm]{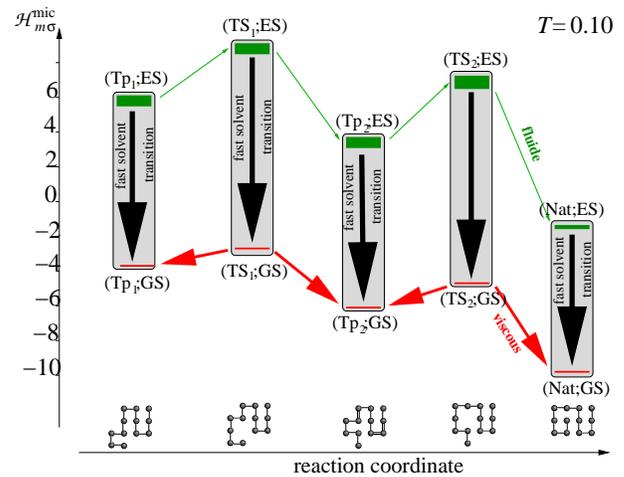}}
\caption{\label{folds10}
The direction and magnitudes of the flux of the connexion of the pathway shown in 
fig.\ref{example} at $t_0$ and $T=0.10$. 
The larger the flux, the wider is the line. Connections with tiny flux are not drawn.}
\end{figure}

At higher temperature, the probability of occurrence of Nat at $t_0$ becomes
higher and the length of the plateau smaller. Many chains in the Tp$_1$,
TS$_1$ or Tp$_2$ jump to Nat.
At $T=0.40$, the probability of the ES becomes very high ($\approx 0.3$) at
a time denoted by $t_M$ in the early events and 
the probability of Nat equals 0.8 very fast.
Between $t_M$ and $t^*$,
the solvent of the chains in ES relax in GS. 
Here, the ES of Nat acts as a strong dynamical attractor.


\ The instantaneous flux from $(m' \sigma')$ to $(m \sigma)$,
given by 
$k_{m' \sigma' \rightarrow m \sigma} = Y_{m\sigma,m'\sigma'} {\cal P}_{m' \sigma'}(t)
- Y_{m'\sigma', m \sigma} {\cal P}_{m\sigma}(t)$ can also be calculated.
Figure \ref{folds10} shows that, at $T=0.10$,
the kinetics is only guided by the difference of energy of the micro-states.

\begin{figure}[htbp]
\centerline{\includegraphics[width=8.0cm]{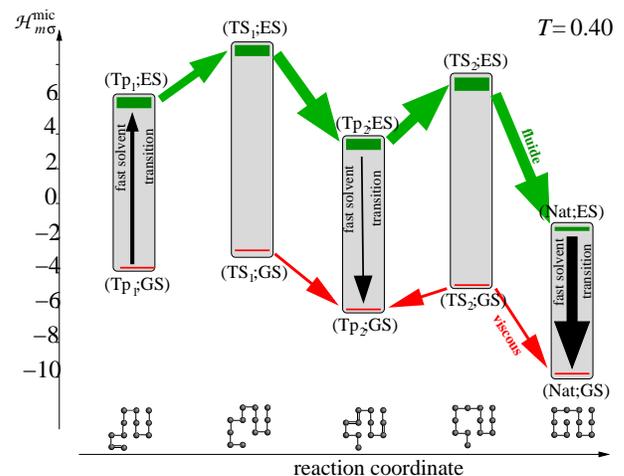}}
\caption{\label{folds40a}
 Same than fig.\ref{folds10} at $T=0.40$. Flux in the GS pathway is the same
for both simulations.}
\end{figure}

The solvated chains have a very large probability
to move towards a micro-state of lower energy.
Between the initial time and $t_0$, the ES of Nat relax very fast to the GS of Nat.
During this period, the ES of all structures relax to the GS. 
Between the times $t_0$ and $t^*$, TS$_1$ and TS$_2$ relax to Tp$_1$, Tp$_2$ and Nat.
The probabilities of GS of Tp$_1$, Tp$_2$ and Nat
increase until all other state have a infinitesimal probability and 
the probability of the Nat increases to one half of that of TS$_2$.
Then, all the fluxes become tiny and the dynamics is frozen.

At $T=0.40$, the equilibrium probabilities of the ES are infinitesimal.
Then, we could also expect a kinetic which would only pass through 
the ground states, as for $T=0.10$. Instead  of this, the sampling of chains
finds another strategy to  overcome the barrier very fast.
At time $t_0$, the probability of the GS of Nat is already of 0.6.
The flux between the pairs of states, drawn in fig.\ref{folds40a}, show that 
the chains in GS of Tp$_1$ do not fold via the GS of the other structures
of the pathway. First, they reach the ES of Tp$_1$ and then they pass through
the ES of the other structures and lastly they relax to the GS of Nat.
This is a consequence of the larger transition rates between the ES than between
the GS. In addition, the probabilities of the ES remain very small.

\vskip0.3cm
{\it 
The chains follow a pathway with large transition rates between improbable states
for which the in going flux equals the out going keeping their probabilities
small.}
\vskip0.3cm

At time $t_0$, the probability of Nat is around 0.7.
The probabilities of the ES are quasi-null.
The flux via the ES pathway is the same as via the GS pathway.
The kinetics become very slow and then a plateau occurs in the curves
of the probability of Nat as a function of the time shown previously.

\section{How water lubricates or freezes the folding. A physical picture.}

\begin{figure}[hbt]
\centerline{\includegraphics[width=8.2cm]{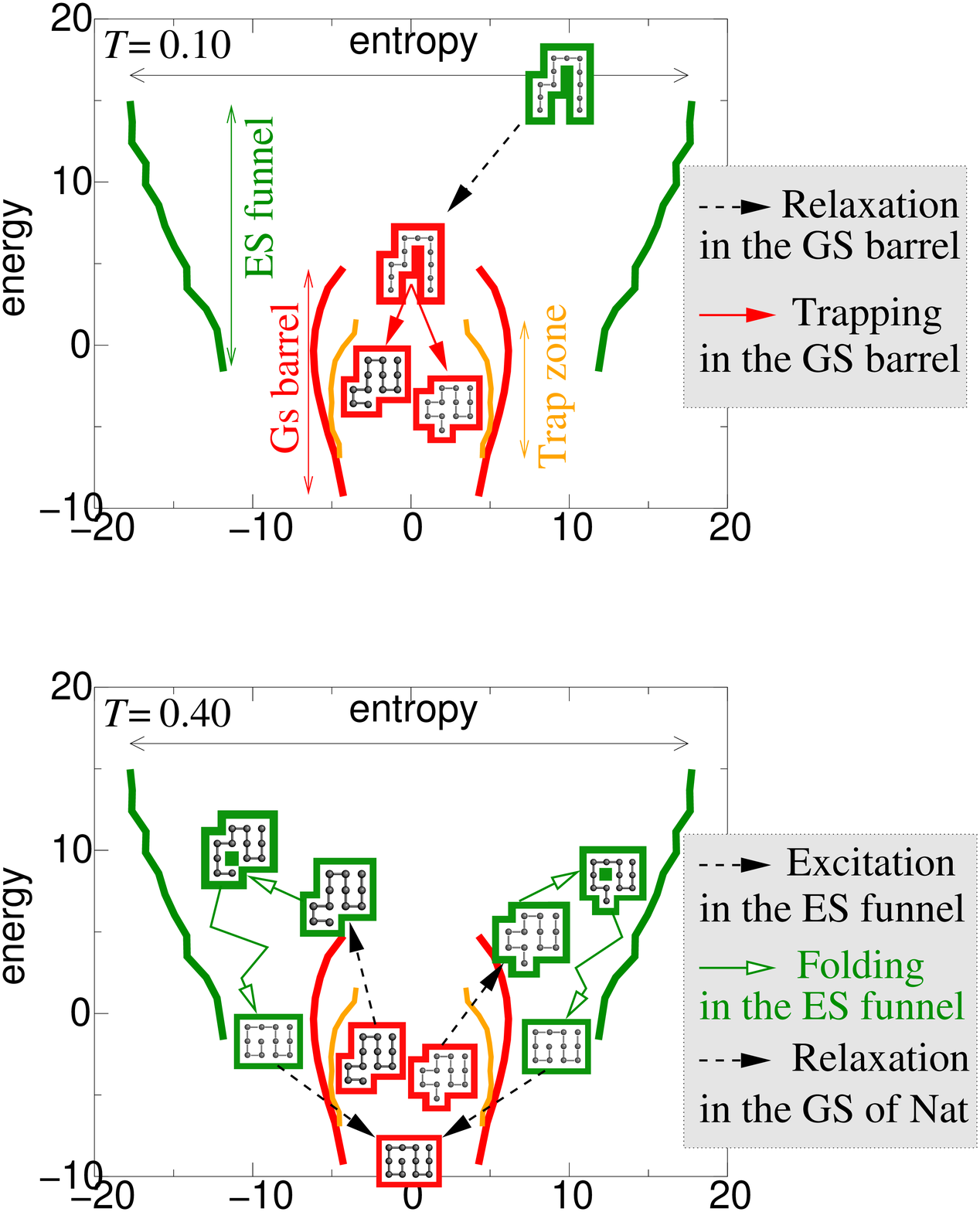}}
\caption{\label{cmp_funnel}
Top: At $T=0.10$, the chains in any energy level of the ES-funnel relax very fast 
in the GS-barrel.
Then, they move down slowly in the barrel, they reach the trap zone and the
dynamics is frozen. 
Only a tiny fraction of the chains, synthesized in chain structures very close to Nat,
reach it in a reasonable time.
Bottom: The picture of the folding at higher temperature but below $T_0$ where
the ES of the chains still has an infinitesimal equilibrium probability, 
depicts a different mechanism.
At $T=0.40$, the chains in the GS-barrel move to the ES-funnel.
Then, most of them fold very fast towards the bottom of the ES-funnel (in the ES of the
native structure)
and they relax in the bottom of the GS barrel (in the GS of the native structure).
They get round the trap zone by passing in the ES-funnel where the transition rates
are very high.
A few chains relax in the GS-barrel during their descent in the ES-funnel and
then they have slow dynamics.
The temperature where the folding mechanism switches from one scenario
to the other one is the glass temperature of the system.
}
\end{figure}

Curves of figs. \ref{main_result_kinetic} and \ref{evolution} are not exactly the
same but are similar.
The basic mechanisms found for the direct fold in the case of the
five-chain conformations pathway
may be extended to the whole configurational space of the protein.
Put together, the results of this paper allow us to give the picture of 
the folding given in fig.\ref{cmp_funnel}.
To draw one of the envelop in the energy-entropy plot,
the entropy, noted $S_\sigma(E)$ for $\sigma=0$, associated to a given microscopic energy, 
is calculated from the number of chain and solvent configurations, 
with formed hydrogen bonds.
The same calculation, of the entropy noted $S_\sigma(E)$ for $\sigma=1$,
is done for the solvent without hydrogen bonds.
The two functions $S_\sigma(E)$ are related to the total number of
protein-solvent configurations whose total energy matches the energy values $E$:
$$S_\sigma(E)  =  \ln [ \sum_m \sum_{\alpha,\beta} \delta(\sigma-\sigma(\beta))
		\ \ \delta^{(\varepsilon)} (E - \Hmic) ] $$
where $\delta^{(\varepsilon)}(x) = 1$ if $-\varepsilon / 2 < x < \varepsilon / 2$
and 0 otherwise.
They may be written as functions of the degeneracy of the macro-states: 
$$S_\sigma(E)  =	  \ln [ \sum_m g_{m \sigma} \ \delta^{(\varepsilon)} (E - \Hsig) ]$$

Moreover a parameter $\theta_m$ allows one to distinguish between the structures of
the bottom of the configurational valleys,
  which may be kinetics traps in the folding.
One defines $\theta_m=1$ for the macro-states, 
only connected to macro-states of higher energies and 0 otherwise. 
Although the native conformation satisfies to this definition, it can not be considered
as a trap and the value of $\theta_{\rm Nat}$ is set to 0.
Considering the five structures of fig. \ref{example}, \ref{folds10} and \ref{folds40a}
as an example, one has $\theta_{\rm Tp_1} = \theta_{\rm Tp_2} = 1$ and
$\theta_{\rm TS_1} = \theta_{\rm TS_2} = \theta_{\rm Nat} = 0$.
Thus, the envelop of the trap region in the energy-entropy plot is a subset 
of the GS barrel, given by~:
$$S_{\rm {Tp}}(E) = \ln [ \sum_m \ \theta_m
g_{m;0}  \ \delta^{(\varepsilon)} (E - {\cal H}^{\rm mac}_{m;0}) ]$$
where $g_{m;0}$ and ${\cal H}^{\rm mac}_{m;0}$ are for the degeneracy and energy of the GS of the structure $m$.

Figure \ref{cmp_funnel} shows the three surfaces drawn on the same plot.
The GS-barrel (respectively ES-funnel) is  associated to the 
GS (respectively ES) of the chain structures.
A part of the surface of the GS-barrel is populated with the
trap conformations 
We have already shown, in this paper, that the connections between two chain structures in the ES-funnel have small
relaxation times and that in the GS-barrel long relaxation times.
This results from the following mechanism.
A single protein and its solvent is in a given configuration, at a given time,
which belongs to a corresponding macro-state, and not in all the configurations
of a the macro-state. 
As a consequence, the transition  
rates between one protein-solvent configuration 
of the macro-state $(m \sigma)$ and those
of a macro-state $(m' \sigma')$ depends on the energy difference between 
these states and also on the degeneracy of $(m' \sigma')$ but not on the
degeneracy of $(m \sigma)$.

In particular,
the connexion from a given configuration of the GS to those of the ES of the same chain 
conformation involves a huge number of routes which increase the energy 
On the opposite, a few connexions allow the backward transition which decreases 
the energy.

{\it
\vskip0.3cm
\noindent At "very low" temperature, the proteins and solvent follow pathways which minimize the microscopic energy.
Then, they very slowly evolve, along the few routes of the GS barrel and fall rapidly in some trap conformations.

\vskip0.3cm
\noindent At "not too low" temperature, proteins and solvent follow  
pathways where the number of routes is maximized.
Then, they quickly evolve along the vast possibilities of routes of
the ES funnel without traps and reach very fast the native structure.
\vskip0.3cm
}

As it has previously been shown, 
the temperature of glass transition which separates the 
two mechanisms is not clearly defined because it depends on the ratio of folding proteins

In addition, we mention that, the calculation of the partition  function
is independent on the informations concerning the network of connexions.
As a consequence, the glass temperature is not related to the temperatures 
($T_0$, $T^*$ and $T_m$) defined above. 
This explains why the excited states of the first shell lubricate the folding under
conditions where only the ground states have non-nil equilibrium probabilities
and why the folding is frozen at lower temperature.

\section{Conclusion.}

We have shown that a model of protein-solvent which takes into account the difference of degeneracies of the bulk solvent and 
the first shell solvent with mainly formed or mainly broken hydrogen bonds 
permits an understanding
of the mechanism which may lead to quasi-instantaneous folding of a sufficiently significant
ratio of the proteins in solution.
Figure \ref{cmp_funnel},
shows that two distinct folding mechanisms exist.
In the first one, the folding times are very large and in the second
one very short.

\begin{acknowledgments}
Acknowledgment to Christophe Chatelain and Bertrand Berche for helpful discussions and to
Rosemary Harris and Chris Chipot for critical reading of the manuscript.
\end{acknowledgments}

\vskip0.5cm
\centerline{\bf Appendix A.}
\vskip0.5cm

\begin{figure}[htbp]
\centerline{\includegraphics[width=5cm]{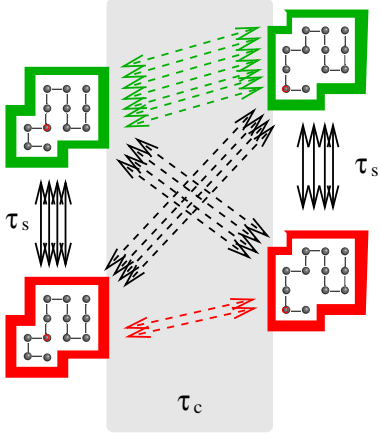}}
\caption{\label{connexion}
Two types of moves, with different characteristic times, are considered in the dynamics:
the protein monomer move (dashed arrows) 
or the solvent configuration transition (other arrows).
For this last event, the connection between solvent configurations of 
the same macroscopic level do not affect the kinetics.
The solvent transition is supposed to have smaller relaxation time than the monomer move.
The relaxation times of the connection are defined as follows:
we solve eqs.\ref{master} and \ref{Xrate} for an isolated connexion
between two states $(m \sigma)$ and $(m' \sigma')$ leading to:
$\Pmic(t) = p^{(\infty)}_{m \alpha \beta} +
[p_{m \alpha \beta}(0) - p^{(\infty)}_{m \alpha \beta}] \exp(-t/\taumic)
$.
$\taumic$ is chosen as $\tau_c$ if the connexion is between two different protein structures
and is chosen as $\tau_s$ if only water moves ($\tau_s \ll \tau_c$).
The set of one monomer moves considered here, is the corner flip (shown in this figure)
and the tail move (not shown).
}
\end{figure}

The probability of occurrence of the conformation 
$m$ and the solvent in micro-state $(\alpha, \ \beta)$ at time $t$ is denoted by
$\Pmic(t)$. 
The master equation of the system is written ~:
\begin{equation}\label{master}
\ds \frac{\d \Pmic} {\d t}  
  = \sum_{m' \alpha' \beta'}   
   \ds X_{m \alpha \beta ;  m' \alpha' \beta'} \ \Pmicp 
\end{equation}
where $X_{m \alpha \beta ;  m' \alpha' \beta'} = X(m' \alpha' \beta' \rightarrow  m \alpha \beta) $ 
is the transition rate from configuration
$(m'\alpha' \beta')$ to $(m  \alpha \beta)$.
The diagonal terms which take into account of the transition from $(m  \alpha \beta)$ to
the other configurations, are~:
$$X_{m \alpha \beta ; m \alpha \beta} = 
-  \sum_{(m' \alpha' \beta') \ne (m \alpha \beta)}
X_{m' \alpha' \beta' ;  m \alpha \beta} 
$$
The detailed balance conditions,
\begin{equation}\label{balanceX}
X_{m \alpha \beta ; m' \alpha' \beta'}  p^{\rm eq}_{m' \alpha' \beta'} =
X_{m' \alpha' \beta' ; m \alpha \beta}  p^{\rm eq}_{m \alpha \beta}
\end{equation}
allow to write the rate of transition as~:
\begin{equation}\label{Xrate}
X_{m \alpha \beta ;  m' \alpha' \beta'}
= \frac{V^{(0)}_{m m'}}{\taumic} \ a_T(\Hmic ; \Hmicp)
\end{equation}
where  $V^{(0)}_{mm'} = V^{(0)}_{m'm} =1$
if the two structures are connected 
by a corner flip and tail moves\cite{Collet2003b} (see fig.\ref{connexion})whatever the solvent configurations
and $0$ otherwise.
The acceptance function is $a_T(x;x') = [1+ \exp((x-x')/T)]^{-1}$ and 
$\taumic$ is a symmetric function~:
$\taumic = \tau_c$ if $m\ne m'$ and $\taumic = \tau_s$ if $m=m'$.

The probability of occurrence of the macro-state $(m \sigma)$ at time $t$ denoted~
$$ \Qmic(t) = \sum_{\alpha, \beta} \Pmic(t) \ \delta(\sigma-\sigma(\beta)) $$
where $\delta(0)=1$ and $\delta(n) = 0$ if $n \ne 0$ and the following relations~: 
$$\begin{array}{ll}
\ds \sum_{\alpha' \beta'} \Pmicp 
& \ds = \sum_{\sigma'} \sum_{\alpha' \beta'} \Pmicp \delta(\sigma'-\sigma(\beta')) \\
& \ds = \sum_{\sigma'} \Q^{\rm mac}_{m \sigma'}
\end{array}$$
$$\sum_{\alpha \beta} \delta(\sigma-\sigma(\beta)) = g_{m \sigma}$$
will be used below.

Now, we rewrite the master equation \ref{master} as~: 
$$	\begin{array}{l}
\ds \sum_{\alpha \beta} \frac{\d \Pmic} {\d t}  \ds \delta(\sigma-\sigma(\beta))
  = \\
\ds \sum_{\alpha \beta} \sum_{m' \alpha' \beta'} \delta(\sigma-\sigma(\beta))
 X_{m \alpha \beta   m' \alpha' \beta'} \ \Pmicp 
\end{array}$$
yields~:
$$ 	\begin{array}{lll}
\ds \sum_{\alpha \beta} \frac{\d \Pmic} {\d t} & \hskip-0.3cm \delta(\sigma-\sigma(\beta))
  = \ds \sum_{m'} \sum_{\alpha \beta}  \delta(\sigma-\sigma(\beta))
	\sum_{\sigma'} \sum_{\alpha' \beta'}  \ \cdot  \\
&  \ds \delta(\sigma'-\sigma(\beta'))
   \frac{V^{(0)}_{mm'}}{\taumic} a_T(\Hmic ; \Hmicp) \ \Pmicp    
\end{array}$$

The evolution equation is rewritten~:
\begin{equation}\label{Ymaster}
\frac{\d \Qmic(t)}{\d t} = \sum_{m'} \sum_{\sigma'} Y_{m \sigma, m' \sigma'} \ \Q^{\rm mac}_{m' \sigma'} 
\end{equation} 
with
\begin{equation}\label{Yrate}
Y_{m \sigma, m' \sigma'} = 
g_{m \sigma}\  \frac{V^{(0)}_{m m'}}{\taumic} \ a_T(\Hsig ; \Hsigp) 
\end{equation} 

In addition, we mention that this result also satisfies the following balance equation:
$Y_{m \sigma, m' \sigma'} {\cal P}^{\rm eq}_{m' \sigma'} =
 Y_{m' \sigma', m \sigma} {\cal P}^{\rm eq}_{m \sigma}$.


\end{document}